# Bidirectional quantum controlled teleportation by using EPR states and entanglement swapping

Shima Hassanpour [1] and Monireh Houshmand [2]

1. Corresponding author, MS Student, Department of Electrical Engineering, Imam Reza International University, Mashhad, Iran.
   shimahassanpour@yahoo.com.
2. Assistant Professor, Department of Electrical Engineering, Imam Reza International University, Mashhad, Iran
   m_houshmand61@yahoo.com.

──**Abstract**──────────────────────────────

In this paper, a novel protocol for bidirectional controlled quantum teleportation (BCQT) is proposed. Based on entanglement swapping of initiate Bell state, two users can teleport an unknown single-qubit state to each other under the permission of the supervisor. This proposed protocol would be utilized to a system in which a controller controls the communication in one direction only. Indeed, just one of the users needs the permission of the controller to reconstruct the unknown quantum state. In comparison to the existing BCQT protocols which their quantum channels are cluster and brown state, the proposed protocol is more practical within today's technology, since it merely uses Bell states as the quantum resource.

## 1 Introduction

Quantum mechanics represents some special capabilities for the transmission of quantum information [1]. Based on the principles of quantum mechanics, there are many forms of quantum communication [2 − 5]. Quantum teleportation (QT) is a type of quantum communication that an unknown quantum state is teleported from one place to the other place via entanglement and with the help of classical information. Many quantum teleportation protocols [6 − 11] have been proposed since Bennett et al., [12] first proposed a QT protocol in 1993.

Controlled quantum teleportation (CQT), first presented by Karlsson and Bourennane [13] in 1998. Over the past few years, much attention has been considered on this interesting topic. Consequently, several CQT protocols by applying various types of entangled state as a quantum channel have been introduced [14 − 19].

In 2013, Zha et al., [20] proposed the first Bidirectional controlled quantum teleportation (BCQT) by employing five-qubit entangled state. After that, based on different types of entangled states as a quantum channel, several BCQT protocols have been suggested [21 − 25]. In all of these BCQT protocols, without the permission of Charlie as a controller the other two users cannot reconstruct an unknown quantum state.

In this paper, a bidirectional controlled quantum teleportation by using Bell-state and applying entanglement swapping technique [26] is proposed. In contrast to the other BCQT protocols [21 − 26], in the proposed scheme, the permission of Charlie as a controller is one direction only. Indeed, just one of the users needs the controller's classical information to reconstruct the quantum state and the other one can make it without the controller's information.

In Section 2, the proposed BCQT protocol based on EPR pairs and entanglement swapping is presented. The comparison between our protocol with previous BCQT protocols is given in Section 3. Finally, the conclusion is presented in Section 4.

## 2 Scheme for bidirectional controlled quantum teleportation

*2.1 Elementary Knowledge*

Before presenting our bidirectional controlled quantum teleportation protocol we need to define

EPR pairs and some unitary operations. An EPR pair can be in one of the four Bell states as described in Eq. (1).

$$|\phi^+\rangle_{12} = \frac{1}{\sqrt{2}}(|00\rangle + |11\rangle)_{12} = \frac{1}{\sqrt{2}}(|++\rangle + |--\rangle)_{12}, \quad |\phi^-\rangle_{12} = \frac{1}{\sqrt{2}}(|00\rangle - |11\rangle)_{12} = \frac{1}{\sqrt{2}}(|+-\rangle + |-+\rangle)_{12},$$

$$|\psi^+\rangle_{12} = \frac{1}{\sqrt{2}}(|01\rangle + |10\rangle)_{12} = \frac{1}{\sqrt{2}}(|++\rangle - |--\rangle)_{12}, \quad |\psi^-\rangle_{12} = \frac{1}{\sqrt{2}}(|01\rangle - |10\rangle)_{12} = \frac{1}{\sqrt{2}}(|+-\rangle - |-+\rangle)_{12},$$

(1)

where $|0\rangle, |1\rangle$ and $|+\rangle, |-\rangle$ are eigenvectors of Pauli operators $\sigma_z$ and $\sigma_x$ respectively.

Now let us describe the four unitary operations. These operations can transform one of the EPR states into another.

$$I = |0\rangle\langle 0| + |1\rangle\langle 1|, \qquad \sigma_x = |0\rangle\langle 1| + |1\rangle\langle 0|,$$

$$i\sigma_y = |0\rangle\langle 1| - |1\rangle\langle 0|, \qquad \sigma_z = |0\rangle\langle 0| - |1\rangle\langle 1|. \tag{2}$$

Also, Hadamard operation is defined as follows,

$$H = \frac{1}{\sqrt{2}}(|0\rangle\langle 0| - |1\rangle\langle 1| + |0\rangle\langle 1| + |1\rangle\langle 0|). \tag{3}$$

*2.2 Description of the proposed scheme*

In this protocol, Alice and Bob as a two legitimate users want to teleport a single qubit state to each other under the permission of the controller. In this scheme, just one of the users requires the controller's classical information to get the unknown quantum state. Suppose Alice and Bob have a single qubit state, which are described as Eq. (4).

$$|\phi\rangle_A = \alpha_0|0\rangle + \alpha_1|1\rangle,$$
$$|\phi\rangle_B = \beta_0|0\rangle + \beta_1|1\rangle, \tag{4}$$

where $|\alpha_0|^2 + |\alpha_1|^2 = 1$ and $|\beta_0|^2 + |\beta_1|^2 = 1$. This protocol consists of the following steps:

Step1. Assume that the quantum channel linking Alice, Bob and Charlie is composed of three EPR entangled state, which has the form of Eq. (5).

$$|\varphi\rangle_{a_1 b_1 c_1 a_2 c_2 b_2} = \frac{1}{\sqrt{2}}(|00\rangle + |11\rangle) \otimes \frac{1}{\sqrt{2}}(|00\rangle + |11\rangle) \otimes \frac{1}{\sqrt{2}}(|00\rangle + |11\rangle)$$

$$= \frac{1}{2\sqrt{2}}|000000\rangle + |000011\rangle + |001100\rangle + |001111\rangle$$

$$+ |110000\rangle + |110011\rangle + |111100\rangle + |111111\rangle)_{a_1 b_1 c_1 a_2 c_2 b_2}, \tag{5}$$

where the qubits $a_1 a_2, b_1 b_2, c_1 c_2$ belong to Alice, Bob and Charlie respectively. The state of the whole system can be expressed as Eq. (6).

$$|\Psi\rangle_{a_1 b_1 c_1 a_2 c_2 b_2 A B} = |G\rangle_{a_1 b_1 c_1 a_2 c_2 b_2} \otimes |\phi\rangle_A \otimes |\phi\rangle_B. \tag{6}$$

Step2. In this step, Alice and Bob make a CNOT operation with qubits $A$ and $B$ as control qubits and qubits $a_1$ and $b_2$ as target respectively. The state will be the form of Eq. (7).

$$|\Psi'\rangle_{a_1 b_1 c_1 a_2 c_2 b_2 A B} = \frac{1}{2\sqrt{2}}[(|000000\rangle + |000011\rangle + |001100\rangle + |001111\rangle$$

$$+ |110000\rangle + |110011\rangle + |111100\rangle + |111111\rangle)_{a_1 b_1 c_1 a_2 c_2 b_2})\alpha_0\beta_0|00\rangle_{AB}$$

$$+ (|000001\rangle + |000010\rangle + |001101\rangle + |001110\rangle$$

$$+ |110001\rangle + |110010\rangle + |111101\rangle + |111110\rangle)_{a_1 b_1 c_1 a_2 c_2 b_2})\alpha_0\beta_1|01\rangle_{AB}$$

$$+ (|100000\rangle + |100011\rangle + |101100\rangle + |101111\rangle$$

$$+ |010000\rangle + |010011\rangle + |011100\rangle + |011111\rangle)_{a_1 b_1 c_1 a_2 c_2 b_2})\alpha_1\beta_0|10\rangle_{AB}$$

$$+ (|100001\rangle + |100010\rangle + |101101\rangle + |101110\rangle$$

$$+ |010001\rangle + |010010\rangle + |011101\rangle + |011110\rangle)_{a_1 b_1 c_1 a_2 c_2 b_2})\alpha_1\beta_1|11\rangle_{AB}. \tag{7}$$

Step3. Alice and Bob perform a single qubit measurement in the $Z$-basis on qubits $a_1$ and $b_2$ and the $X$-basis measurement on qubits $A$ and $B$ respectively. According to Table I, the remaining particles may collapse into one of the 16 possible state with the same probability.

**TABLE I.** THE MEASUREMENT RESULTS OF USERS AND THE CORRESPONDING COLLAPSED STATE

| Alice's results | Bob's results | The collapsed state of qubits $b_1 c_1 a_2 c_2$ |
|---|---|---|
| $\|0\rangle_{a_1}\|+\rangle_A$ | $\|0\rangle_{b_2}\|+\rangle_B$ | $\alpha_0\beta_0(\|0000\rangle + \|0110\rangle)_{b_1c_1a_2c_2} + \alpha_0\beta_1(\|0001\rangle + \|0111\rangle)_{b_1c_1a_2c_2}$ $+ \alpha_1\beta_0(\|1000\rangle + \|1110\rangle)_{b_1c_1a_2c_2} + \alpha_0\beta_0(\|1001\rangle + \|1111\rangle)_{b_1c_1a_2c_2}$ |
| $\|0\rangle_{a_1}\|+\rangle_A$ | $\|0\rangle_{b_2}\|-\rangle_B$ | $\alpha_0\beta_0(\|0000\rangle + \|0110\rangle)_{b_1c_1a_2c_2} - \alpha_0\beta_1(\|0001\rangle + \|0111\rangle)_{b_1c_1a_2c_2}$ $+ \alpha_1\beta_0(\|1000\rangle + \|1110\rangle)_{b_1c_1a_2c_2} - \alpha_1\beta_1(\|1001\rangle + \|1111\rangle)_{b_1c_1a_2c_2}$ |
| $\|0\rangle_{a_1}\|-\rangle_A$ | $\|0\rangle_{b_2}\|+\rangle_B$ | $\alpha_0\beta_0(\|0000\rangle + \|0110\rangle)_{b_1c_1a_2c_2} + \alpha_0\beta_1(\|0001\rangle + \|0111\rangle)_{b_1c_1a_2c_2}$ $- \alpha_1\beta_0(\|1000\rangle + \|1110\rangle)_{b_1c_1a_2c_2} - \alpha_1\beta_1(\|1001\rangle + \|1111\rangle)_{b_1c_1a_2c_2}$ |
| $\|0\rangle_{a_1}\|-\rangle_A$ | $\|0\rangle_{b_2}\|-\rangle_B$ | $\alpha_0\beta_0(\|0000\rangle + \|0110\rangle)_{b_1c_1a_2c_2} - \alpha_0\beta_1(\|0001\rangle + \|0111\rangle)_{b_1c_1a_2c_2}$ $- \alpha_1\beta_0(\|1000\rangle + \|1110\rangle)_{b_1c_1a_2c_2} + \alpha_1\beta_1(\|1001\rangle + \|1111\rangle)_{b_1c_1a_2c_2}$ |
| $\|0\rangle_{a_1}\|+\rangle_A$ | $\|1\rangle_{b_2}\|+\rangle_B$ | $\alpha_0\beta_0(\|0001\rangle + \|0111\rangle)_{b_1c_1a_2c_2} + \alpha_0\beta_1(\|0000\rangle + \|0110\rangle)_{b_1c_1a_2c_2}$ $+ \alpha_1\beta_0(\|1001\rangle + \|1111\rangle)_{b_1c_1a_2c_2} + \alpha_1\beta_1(\|1000\rangle + \|1110\rangle)_{b_1c_1a_2c_2}$ |
| $\|0\rangle_{a_1}\|+\rangle_A$ | $\|1\rangle_{b_2}\|-\rangle_B$ | $\alpha_0\beta_0(\|0001\rangle + \|0111\rangle)_{b_1c_1a_2c_2} - \alpha_0\beta_0(\|0000\rangle + \|0110\rangle)_{b_1c_1a_2c_2}$ $+ \alpha_0\beta_0(\|1001\rangle + \|1111\rangle)_{b_1c_1a_2c_2} - \alpha_0\beta_0(\|1000\rangle + \|1110\rangle)_{b_1c_1a_2c_2}$ |
| $\|0\rangle_{a_1}\|-\rangle_A$ | $\|1\rangle_{b_2}\|+\rangle_B$ | $\alpha_0\beta_0(\|0001\rangle + \|0111\rangle)_{b_1c_1a_2c_2} + \alpha_0\beta_1(\|0000\rangle + \|0110\rangle)_{b_1c_1a_2c_2}$ $- \alpha_1\beta_0(\|1001\rangle + \|1111\rangle)_{b_1c_1a_2c_2} - \alpha_1\beta_1(\|1000\rangle + \|1110\rangle)_{b_1c_1a_2c_2}$ |
| $\|0\rangle_{a_1}\|-\rangle_A$ | $\|1\rangle_{b_2}\|-\rangle_B$ | $\alpha_0\beta_0(\|0001\rangle + \|0111\rangle)_{b_1c_1a_2c_2} - \alpha_0\beta_1(\|0000\rangle + \|0110\rangle)_{b_1c_1a_2c_2}$ $- \alpha_1\beta_0(\|1001\rangle + \|1111\rangle)_{b_1c_1a_2c_2} + \alpha_1\beta_1(\|1000\rangle + \|1110\rangle)_{b_1c_1a_2c_2}$ |
| $\|1\rangle_{a_1}\|+\rangle_A$ | $\|0\rangle_{b_2}\|+\rangle_B$ | $\alpha_0\beta_0(\|1000\rangle + \|1110\rangle)_{b_1c_1a_2c_2} + \alpha_0\beta_1(\|1001\rangle + \|1111\rangle)_{b_1c_1a_2c_2}$ $+ \alpha_1\beta_0(\|0001\rangle + \|0111\rangle)_{b_1c_1a_2c_2} + \alpha_1\beta_1(\|0000\rangle + \|0110\rangle)_{b_1c_1a_2c_2}$ |
| $\|1\rangle_{a_1}\|+\rangle_A$ | $\|0\rangle_{b_2}\|-\rangle_B$ | $\alpha_0\beta_0(\|1000\rangle + \|1110\rangle)_{b_1c_1a_2c_2} - \alpha_0\beta_1(\|1001\rangle + \|1111\rangle)_{b_1c_1a_2c_2}$ $+ \alpha_1\beta_0(\|0001\rangle + \|0111\rangle)_{b_1c_1a_2c_2} - \alpha_1\beta_1(\|0000\rangle + \|0110\rangle)_{b_1c_1a_2c_2}$ |
| $\|1\rangle_{a_1}\|-\rangle_A$ | $\|0\rangle_{b_2}\|+\rangle_B$ | $\alpha_0\beta_0(\|1000\rangle + \|1110\rangle)_{b_1c_1a_2c_2} + \alpha_0\beta_1(\|1001\rangle + \|1111\rangle)_{b_1c_1a_2c_2}$ $- \alpha_1\beta_0(\|0001\rangle + \|0111\rangle)_{b_1c_1a_2c_2} - \alpha_1\beta_1(\|0000\rangle + \|0110\rangle)_{b_1c_1a_2c_2}$ |
| $\|1\rangle_{a_1}\|-\rangle_A$ | $\|0\rangle_{b_2}\|-\rangle_B$ | $\alpha_0\beta_0(\|1000\rangle + \|1110\rangle)_{b_1c_1a_2c_2} - \alpha_0\beta_1(\|1001\rangle + \|1111\rangle)_{b_1c_1a_2c_2}$ $- \alpha_1\beta_0(\|0001\rangle + \|0111\rangle)_{b_1c_1a_2c_2} + \alpha_1\beta_1(\|0000\rangle + \|0110\rangle)_{b_1c_1a_2c_2}$ |
| $\|1\rangle_{a_1}\|+\rangle_A$ | $\|1\rangle_{b_2}\|+\rangle_B$ | $\alpha_0\beta_0(\|1001\rangle + \|1111\rangle)_{b_1c_1a_2c_2} + \alpha_0\beta_1(\|1000\rangle + \|1110\rangle)_{b_1c_1a_2c_2}$ $+ \alpha_1\beta_0(\|0001\rangle + \|0111\rangle)_{b_1c_1a_2c_2} + \alpha_1\beta_1(\|0000\rangle + \|0110\rangle)_{b_1c_1a_2c_2}$ |
| $\|1\rangle_{a_1}\|+\rangle_A$ | $\|1\rangle_{b_2}\|-\rangle_B$ | $\alpha_0\beta_0(\|1001\rangle + \|1111\rangle)_{b_1c_1a_2c_2} - \alpha_0\beta_1(\|1000\rangle + \|1110\rangle)_{b_1c_1a_2c_2}$ $+ \alpha_1\beta_0(\|0001\rangle + \|0111\rangle)_{b_1c_1a_2c_2} - \alpha_1\beta_1(\|0000\rangle + \|0110\rangle)_{b_1c_1a_2c_2}$ |
| $\|1\rangle_{a_1}\|-\rangle_A$ | $\|1\rangle_{b_2}\|+\rangle_B$ | $\alpha_0\beta_0(\|1001\rangle + \|1111\rangle)_{b_1c_1a_2c_2} + \alpha_0\beta_1(\|1000\rangle + \|1110\rangle)_{b_1c_1a_2c_2}$ $- \alpha_1\beta_0(\|0001\rangle + \|0111\rangle)_{b_1c_1a_2c_2} - \alpha_1\beta_1(\|0000\rangle + \|0110\rangle)_{b_1c_1a_2c_2}$ |
| $\|1\rangle_{a_1}\|-\rangle_A$ | $\|1\rangle_{b_2}\|-\rangle_B$ | $\alpha_0\beta_0(\|1001\rangle + \|1111\rangle)_{b_1c_1a_2c_2} - \alpha_0\beta_1(\|1000\rangle + \|1110\rangle)_{b_1c_1a_2c_2}$ $- \alpha_1\beta_0(\|0001\rangle + \|0111\rangle)_{b_1c_1a_2c_2} + \alpha_1\beta_1(\|0000\rangle + \|0110\rangle)_{b_1c_1a_2c_2}$ |

Step4. After Alice (Bob) tells the result to Bob (Alice) and Charlie, if Charlie wants to co-operate with the other two users, he applies Hadamard operation on his two qubits. As an example, if Alice's and Bob's measurement results in the first step is $|0\rangle_{a_1}|+\rangle_A$, and $|0\rangle_{b_2}|+\rangle_B$, the state of the remaining particles collapse into the state as Eq. (8) shows.

$$|\Omega\rangle_{b_1c_1a_2c_2} = \frac{1}{4\sqrt{2}}[\alpha_0\beta_0(|0000\rangle + |0001\rangle + |0100\rangle + |0101\rangle + |0010\rangle + |0011\rangle - |0110\rangle - |0111\rangle)_{b_1c_1a_2c_2}$$
$$+\alpha_0\beta_1(|0000\rangle - |0001\rangle + |0100\rangle - |0101\rangle + |0010\rangle - |0011\rangle - |0110\rangle + |0111\rangle)_{b_1c_1a_2c_2}$$
$$+\alpha_1\beta_0(|1000\rangle + |1001\rangle + |1100\rangle + |1101\rangle + |1010\rangle + |1011\rangle - |1110\rangle - |1111\rangle)_{b_1c_1a_2c_2}$$
$$+\alpha_1\beta_1(|1000\rangle - |1001\rangle + |1100\rangle - |1101\rangle + |1010\rangle - |1011\rangle - |1110\rangle + |1111\rangle)_{b_1c_1a_2c_2}].$$
(8)

Then Charlie performs Bell measurement and announces his result to the users. The state is as follows:

$$|\Omega\rangle_{b_1c_1a_2c_2} = \frac{1}{4}[|\emptyset^+\rangle_{c_1c_2}(\alpha_0|0\rangle + \alpha_1|1\rangle)_{b_1}(\beta_0|0\rangle + \beta_1|1\rangle)_{a_2} + |\emptyset^-\rangle_{c_1c_2}(\alpha_0|0\rangle + \alpha_1|1\rangle)_{b_1}(\beta_0|1\rangle + \beta_1|0\rangle)_{a_2}$$
$$+|\Psi^+\rangle_{c_1c_2}(\alpha_0|0\rangle + \alpha_1|1\rangle)_{b_1}(\beta_0|0\rangle - \beta_1|1\rangle)_{a_2} + |\Psi^-\rangle_{c_1c_2}(\alpha_0|0\rangle + \alpha_1|1\rangle)_{b_1}(\beta_0|1\rangle - \beta_1|0\rangle)_{a_2}].$$
(9)

Now, each legitimate user can reconstruct the unknown single-qubit state by applying suitable unitary operation as we can see in Table II. Thus, the bidirectional controlled teleportation is successfully finished.

**TABLE II.** RELATION BETWEEN THE MEASUREMENT RESULTS AND APPROPRIATE UNITARY OPERATION

| Charlie's measurement result | $|\emptyset^+\rangle$ | $|\emptyset^-\rangle$ | $|\Psi^+\rangle$ | $|\Psi^-\rangle$ |
|---|---|---|---|---|
| Bob's operation | $I_{b_1}$ | $\sigma^x_{b_1}$ | $\sigma^z_{b_1}$ | $\sigma^{iy}_{b_1}$ |
| Alice's operation | $I_{a_2}$ | $I_{a_2}$ | $I_{a_2}$ | $I_{a_2}$ |

## 3 Comparison

Different from the previous BCQT protocols [20 − 25], the proposed protocol can be applied to a system in which the third party as a controller controls the communication in one direction only. Actually, just one of the users requires the controller's classical information to reconstruct the unknown quantum state.

Unlike other schemes which their quantum channels are cluster and brown state, this scheme utilizes entanglement swapping technique for sharing three-EPR pair. Therefore, the proposed protocol is experimentally more efficient compared with related works in two aspects. On the one hand, the quantum channel is easier to be prepared [27]; on the other hand, the entanglement must be held between two qubits, while preserving entanglement between more than two qubits is more complicated.

Also, in the proposed scheme and the other bidirectional controlled quantum teleportation protocols, legitimate users can teleport an unknown single-qubit state to each other simultaneously. Table III gives us more details of this comparison. In Table III, T.D and O.D refer to types of Charlie's control which denote two-direction and one-direction respectively. In addition, B.M and S.P.M indicates the Bell measurement and the single photon measurement respectively.

TABLE III. THE COMPARISON BETWEEN PROPOSED BCQT PROTOCOL WITH EXISTING ONES

| Ref. | [20] | [21] | [22] | [23] | [24] | [25] | Proposed protocol |
|---|---|---|---|---|---|---|---|
| Type of channel | $Cluster_5$ | $Cluster_6$ | Six-qubit | $Brown_5$ | Seven-qubit | Six-qubit | $EPR_6$ |
| Type of control | T.D | T.D | T.D | T.D | T.D | T.D | O.D |
| Measurement method | S.P.M | S.P.M | B.M | B.M | B.M | B.M | B.M |

# 4 Conclusion

In summary, we have proposed a novel bidirectional controlled quantum teleportation scheme based on EPR pairs and entanglement swapping. In the proposed protocol, Alice and Bob can teleport an unknown single-qubit state to each other simultaneously and Charlie controls the communication. In this scheme, the control of Charlie is one direction. Indeed, just Bob can reconstruct the unknown single-qubit state with the permission from Charlie.

It is worth to be mentioned that, the quantum channel in the proposed scheme is composed of three-EPR pair. Therefore, in the future, it will be implemented in experiment more easily than other BCQT protocols.